\documentclass[aps,pra,longbibliography,twocolumn,superscriptaddress,
floatfix]{revtex4-1}
\usepackage{amsmath}
\usepackage[next]{inputenc}
\usepackage[dvips]{epsfig}
\usepackage{bbm,bm,bbold}
\usepackage{booktabs} 
\usepackage{svg} 
\usepackage{multirow}
\usepackage{hhline}
\usepackage{comment}
\usepackage{braket}
\usepackage{float}
\usepackage{stackengine}
\usepackage{amsmath,amsfonts,amssymb,amsthm}
\usepackage{color}
\usepackage{xcolor}
\usepackage{tikz-feynman}
\usepackage{graphicx,latexsym}

\usepackage[colorlinks=true,urlcolor=blue,citecolor=blue,linkcolor=blue]{hyperref}

\usepackage{lipsum}

\usepackage{nameref}
\usepackage{varioref}
\usepackage{cleveref}

\usepackage{natbib}

\begin{document}
\renewcommand{\vec}{\mathbf}
\renewcommand{\Re}{\mathop{\mathrm{Re}}\nolimits}
\newcommand{\para}{\parallel}
\renewcommand{\Im}{\mathop{\mathrm{Im}}\nolimits}
\newcommand\scalemath[2]{\scalebox{#1}{\mbox{\ensuremath{\displaystyle #2}}}}

\title{Topological hybridisation of plasmons with
ferrimagnetic magnons}

\author{Cooper Finnigan}
\email{cooper.finnigan@monash.edu}
\affiliation{School of Physics and Astronomy, Monash University, Victoria 3800, Australia}

\author{Mehdi Kargarian}
\affiliation{Department of Physics, Sharif University of Technology, Tehran 14588-89694, Iran}

\author{Dmitry K. Efimkin}
\email{dmitry.efimkin@monash.edu}
\affiliation{School of Physics and Astronomy, Monash University, Victoria 3800, Australia}

\begin{abstract}
We study the formation of hybrid plasmon-magnon modes in a heterostructure comprising a monolayer semiconductor with strong Rashba spin-orbit coupling --- specifically, Janus transition-metal dichalcogenides (TMDs) ---  and an insulating ferrimagnet, such as yttrium iron garnet-based compounds. Using a combined microscopic-macroscopic framework for plasmon-magnon coupling, we show that plasmons and magnons strongly hybridize over both GHz and THz frequency ranges, enabling experimental access well above cryogenic temperatures. Moreover, the developed approach provides an efficient and natural classification of the topology of the hybrid modes, rooted in the phase winding of the plasmon-magnon coupling induced by spin-momentum locking and the associated chiral winding of the electronic spin along the Fermi contours. Finally, we identify experimentally accessible manifestations of the hybridization, such as topological interface modes and an anomalous thermal Hall response.
\end{abstract}

\date{\today}
\maketitle
\section{Introduction}
Over the last decade, magnetic heterostructures have attracted intense and growing interest as a platform for both fundamental studies and spintronic applications~\cite{kimFerrimagneticSpintronics2022a, ReviewSpintronicsApp1, bhattacharyyaRecentProgressProximity2021}. By interfacing magnetic order with electronic or collective degrees of freedom, these systems host a remarkably broad range of phenomena, including spin transport and spin-transfer torques~\cite{STTReview}, spin Hall magnetoresistance~\cite{SpinHallMagnetoresistance1}, and anomalous Hall effects up to their quantized limit~\cite{QAHE}. More recently, magnetic heterostructures have also been explored as a setting for emergent collective behavior, such as magnon-mediated superconductivity~\cite{SC1,SC2,SC3,SC4,SC5} and superconducting spintronics~\cite{LinderJacob2015Ss,SCSpintronicsReview2}.

Recently, magnetic heterostructures have emerged as a promising platform for engineering strong coupling between magnons and plasmons~\cite{efimkinTopologicalSpinplasmaWaves2021a, dyrdalMagnonplasmonHybridizationMediated2023, bludovHybridPlasmonmagnonPolaritons2019, ghoshPlasmonmagnonInteractionsTwodimensional2023,PlasmonMagnonNew1,hirosawa2025topologicalmagnonplasmonhybrids, gunnink2026couplingplasmonstwomagnoncontinuum} or surface plasmon-polaritons~\cite{SurfacePlasmonNew1,SurfacePlasmonNew2, SurfacePlasmonNew3,SurfacePlasmonNew4}, and the resulting hybrid modes can be classified as topologically nontrivial~\cite{efimkinTopologicalSpinplasmaWaves2021a}.  This development is motivated more broadly by the growing recognition of, and interest in, beyond-electron topological physics, which can also emerge in systems where collective modes of distinct physical origin hybridize. Prominent examples include exciton-polaritons \cite{PhysRevX.5.031001}, magnon-phonon hybrid modes \cite{goTopologicalMagnonPhononHybrid2019,MagnonPhononNew1,MagnonPhononNew2}, Weyl helicon-phonon modes~\cite{EfimkinHelicon}, and engineered phonon-polariton modes in photonic metasurfaces~\cite{TopPhotonPolariton}, where mode mixing gives rise to robust trapped modes and anomalous transport phenomena.

To date, however, experimental realizations of topological plasmon-magnon modes have not yet been reported. One major obstacle is the mismatch of characteristic energy scales: ferromagnetic magnons typically reside in the GHz frequency range, such that achieving strong coupling to plasmons requires cryogenic (often sub-kelvin) temperatures and exceptionally high plasmonic quality factors. While such conditions can be realized in noble metals and graphene, they are considerably more challenging to achieve on the surfaces of topological insulators and in monolayer semiconductors with strong spin-orbit coupling, where plasmon-magnon interactions have been argued to be enhanced~\cite{efimkinTopologicalSpinplasmaWaves2021a, dyrdalMagnonplasmonHybridizationMediated2023}. It is therefore highly desirable to extend strong plasmon-magnon coupling--- and the associated topological phenomena --- into the THz frequency range. In this regime, plasmons in two-dimensional materials have been extensively studied, and the requirements for reaching the strong-coupling limit are substantially more favorable \cite{FeiZ2012Gogp, chenOpticalNanoimagingGatetunable2012a,grigorenkoGraphenePlasmonics2012a,xuTimedomainStudyCoupled2025}.
\begin{figure}[b]
    \vspace{-0.10in}
    \centering
    \includegraphics[width=\columnwidth]{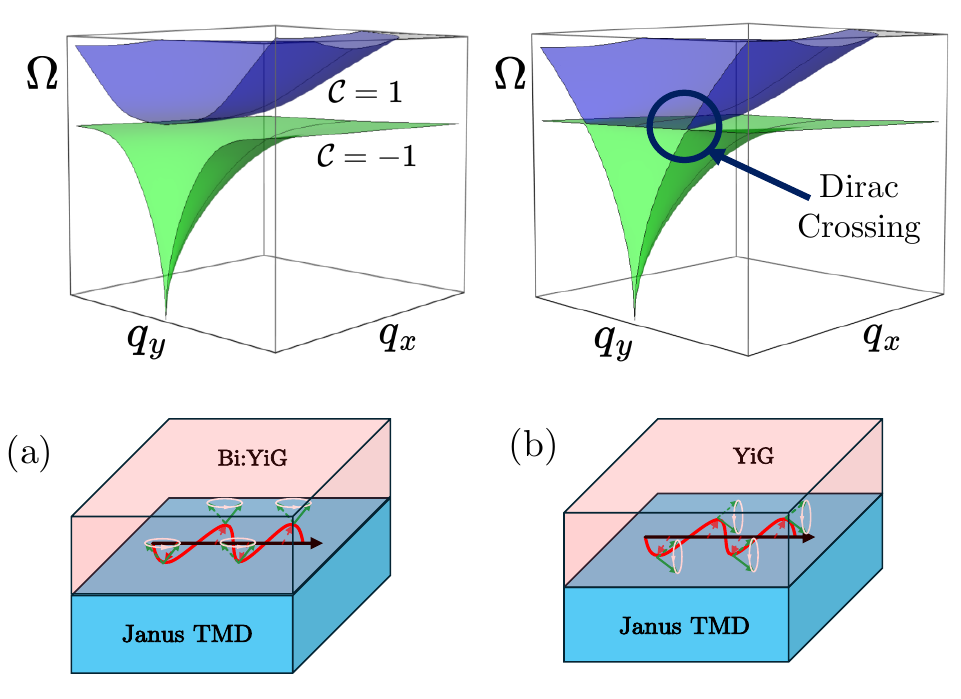}
    \caption{
    Illustration of a Janus TMD interfaced with an insulating ferrimagnet for (a) \textit{out-of-plane} and (b) \textit{in-plane} magnetization, together with sketches of the corresponding hybrid plasmon--magnon spectra. Black (red) arrows denote electron current (spin) oscillations, while precessing green arrows indicate magnetization fluctuations. For out-of-plane magnetization, the spectrum is gapped and isotropic, exhibiting nonzero topological Chern numbers. For in-plane magnetization, the spectrum exhibits a pair of Dirac branch-crossing nodes and is strongly anisotropic. }
\label{Figure_1} 
\end{figure}

In this paper, we investigate the magnetic proximity effect between a Rashba electron gas in a monolayer semiconductor with broken inversion symmetry --- such as a Janus transition-metal dichalcogenides (e.g., MoSeTe or WSeTe) \cite{zhangJanusMonolayerTransitionMetal2017, yaoManipulationLargeRashba2017, REZAVAND2022168721} --- and an insulating ferrimagnet, such as yttrium iron garnet (YIG) \cite{princepFullMagnonSpectrum2017,AnisotropyYIG1} or its Bi-substituted counterpart (Bi:YIG)~\cite{AnisotropyYIG2}. Owing to spin-momentum locking of Rashba electrons, plasmon excitations exhibit coupled charge- and spin-density oscillations and therefore interact efficiently with magnetic fluctuations across the interface.  We show that both magnon branches, carrying opposite helicities, hybridize with plasmons, giving rise to mixed plasmon-magnon modes spanning frequency ranges from the GHz to the THz regime. This behavior reflects the presence of a finite magnetic polarization in ferrimagnets, in contrast to multisublattice ferromagnets and antiferromagnets, where optical magnon modes residing in the THz regime are spin-unpolarized~\cite{kimFerrimagneticSpintronics2022a,fennerNoncollinearitySpinFrustration2009, princepFullMagnonSpectrum2017}. 
For out-of-plane magnetic anisotropy, as realized in Bi:YIG, the hybrid spectrum is fully gapped and all branches acquire nontrivial Chern numbers, signaling a topological phase. By contrast, for in-plane anisotropy, as realized in YIG, the gap closes when plasmon propagation is perpendicular to the magnetic ordering, and the corresponding hybrid modes exhibit Dirac-type crossings. These features are captured within the ``effective Hamiltonian" framework developed here, which combines a macroscopic description of plasmons with microscopic response functions of the Rashba electron gas. Finally, we discuss experimentally accessible signatures of the hybridization, including topological interface states and an anomalous thermal Hall effect.

The remainder of the paper is organized as follows. In Sec.~\ref{SecIIa}, we analyze the proximity effect involving a ferrimagnet with out-of-plane anisotropy, presenting the underlying model as well as both microscopic and macroscopic descriptions of the resulting hybrid plasmon-magnon modes and their topological classification. In Sec.~\ref{SecIIb}, we extend this framework to the case of a ferrimagnet with in-plane anisotropy. Finally, Sec.~\ref{SecIII} contains a discussion of the results and concluding remarks.

\section{hybrid plasmon-magnon modes }
\subsection{Out-of-plane magnetic ordering}
\label{SecIIa}
Consider the interface between a monolayer semiconductor with broken inversion symmetry and an insulating ferrimagnet with out-of-plane magnetisation. The semiconductor hosts electrons with Rashba-type spin-orbit coupling, giving rise to collective plasmonic excitations that manifest as propagating coupled charge- and spin-density waves \cite{wangPlasmonSpectrumTwodimensional2005,PhysRevLett.104.116401,EfimkinDmitryK2012Ceoa,efimkinSpinplasmonsTopologicalInsulator2012}. The ferrimagnet, in turn, supports propagating spin waves, or magnons. Direct plasmon-magnon coupling mediated by an oscillating magnetic field is relativistic in origin and therefore negligibly weak compared to the indirect exchange-mediated coupling arising from interfacial interactions between the electron spins and local magnetic moments in the ferrimagnet. 

For the sake of simplicity, we assume that the ferrimagnet has two sublattices and therefore supports a pair of magnon modes. In contrast to multisublattice ferromagnets and antiferromagnets, both branches $\alpha=\mathrm{R} /  \mathrm{L}$ carry uncompensated oscillating spin-polarisation, $\boldsymbol{\ell}^\alpha(\vec{r},t)$, which is perpendicular to the unit vector $\vec{n}=e_\mathrm{z}$ that aligns the out-of-plane spin ordering. After linearisation, the dynamics of both magnon branches become independent, and are governed by two independent Landau-Lifshitz-Gilbert (LLG) equations:
 \begin{equation}
 \rho_\mathrm{s}\left(\alpha\partial_{t}\left[ \boldsymbol{\ell}^{\alpha}(\mathbf{r}, t)\times \vec{e}_{z}\right]-\omega^{\alpha}_{\boldsymbol{q}}\boldsymbol{\ell}^{\alpha}(\mathbf{r}, t)\right)= \Delta_{\alpha} \vec{s}(\vec{r},t).\label{LLG}
 \end{equation}
Here, $\rho_{\mathrm{S}}$ is the magnetic spin density, and $\omega^{\alpha}_{\boldsymbol{q}}$ is the dispersion of the magnon branches~\footnote{(See Appendix.~\ref{MICMagnons}) for their explicit expressions for a square lattice formed by alternating spins $S_1$ and $S_2$, a  minimal model for ferrimagnet}. The coupling strength $\Delta_{\alpha}$ between magnons and the electronic spin density $\mathbf{s}(\mathbf{r},t)$ is branch dependent, reflecting the different magnitudes and sign of the spin polarization carried by the two branches. As for plasmons, they manifest via oscillations of the scalar potential $\phi(\mathbf{r},t)$ generated by electron charge density oscillations $\rho(\mathbf{r},t)$ as
\begin{equation}
    e \phi(\mathbf{r},t)=\int d \mathbf{r}^{\prime} V(\mathbf{r}-\mathbf{r}^{\prime}) \rho(\mathbf{r},t). \label{Poisson}
\end{equation}
Here, $V(\mathbf{r})=2\pi/\kappa r$ is the electron-electron interaction and $\kappa$ is the effective dielectric constant for the considered setup. Extension to a more complex dielectric or metallic gate-induced screening is straightforward. 

The equations~(\ref{LLG}) and~(\ref{Poisson}), which govern magnons and plasmons, are not directly coupled and remain independent until spin--orbit interactions entrain the charge- and spin-density dynamics of the electron gas in the monolayer semiconductor. Within a mean-field description, electrons interact both with the self-consistent macroscopic potential $\phi(\mathbf{r},t)$ and magnetic fluctuations $\ell^\alpha(\vec{r,t})$, and their single-particle dynamics are described by the following Hamiltonian  
\begin{equation}
\begin{split}
\mathcal{H}=\frac{\mathbf{p}^2}{2 m}+\alpha_{\mathrm{R}}[\mathbf{p} \times \boldsymbol{\sigma}]_z +(\Delta_{\mathrm{R}}-\Delta_{\mathrm{L}}) \boldsymbol{\sigma}\cdot \vec{n} -\epsilon_\mathrm{F}  \\ + 
\sum_\alpha \Delta_{\alpha} \boldsymbol{\sigma}\cdot \boldsymbol{\ell}^{\alpha}(\vec{r,t})+ e \phi(\vec{r},t).
\end{split}
\end{equation}
Here $m$ is the effective mass, $\alpha_{\mathrm{R}}$ is the Rashba coupling, and $\epsilon_\mathrm{F}$ is the Fermi level for electrons. The electron band dispersion of Rashba electrons, $E^{\sigma}(\mathbf{p})= \mathbf{p}^{2}/2m + \sigma\sqrt{\alpha_{\mathrm{R}}^{2}\ p^{2}+(\Delta_{\mathrm{R}}-\Delta_{\mathrm{L}})^{2}}$, has two helicities $\sigma = \uparrow/\downarrow$, with electronic spin textures exhibiting opposite chiral windings. The two Rashba subbands have two different band-dependent Fermi momenta \cite{PhysRevB.91.035106}, given as
\begin{equation}
    p_{\mathrm{F}}^{\sigma} = \sqrt{2m \varepsilon_{\mathrm{F}} + ( m \alpha_{\mathrm{R}})^{2}} - \sigma m \alpha_{\mathrm{R}}.
\end{equation}
Since the two ferrimagnetic modes couple to electrons in a similar way, we will present only the analysis of the coupling with the right-handed magnon mode, but include both chiralities in the plots presented below. 

To obtain a closed set of equations, Eqs.~(\ref{LLG}) and~(\ref{Poisson}) must be supplemented with the matter response. In this work we follow the hydrodynamic scheme to describe plasmons \cite{FETTER1973367,FETTER19741}, excluding electron density using the continuity equation
\begin{equation}
    \partial_{t} \rho(\mathbf{r},t) + \nabla \cdot \mathbf{j}(\mathbf{r},t) = 0, \label{Continuity}
\end{equation}
and calculate the linear response of electron spin $\mathbf{s}(\mathbf{r},t)$ and current $\mathbf{j}(\mathbf{r},t)$ densities using the Kubo theory~\footnote{In our previous work, we formulated the theory describing plasmon-magnon coupling in terms of coupled response functions of charge $\rho(\mathbf{r},t)$ and spin-densities $\mathbf{s}(\mathbf{r},t)$ to the scalar potential $\phi(\mathbf{r},t)$ and magnetic fluctuations $\boldsymbol{\ell}_{\alpha}(\mathbf{r},t)$ evaluated using the Kubo linear-response framework. Although this formulation provides a transparent physical interpretation and enables a straightforward determination of the mixed-mode dispersion, it is not optimal for analysing their topological properties.}. In the local and the high-frequency approximations (See Appendix.~\ref{Ap.1} for details) the response functions can be presented as 
\begin{equation}
\begin{gathered}
\mathbf{j}(\mathbf{r},\omega) = -\frac{e^{2}v_{\mathrm{F}}^{2}N_{\mathrm{F}}^{+}}{i\omega} \mathbf{E}(\mathbf{r},\omega) + e v_{\mathrm{F}}  \Delta_{\alpha}N_{\mathrm{F}}^{\mathrm{-}}\left[\mathbf{e}_{z}\times\boldsymbol{\ell}^{\alpha}(\mathbf{r},\omega)\right],\\
 \boldsymbol{\vec{s}}(\mathbf{r},\omega)=  \Delta_{\alpha} N_{\mathrm{F}}^{+}\boldsymbol{\ell}^{\alpha}(\mathbf{r},\omega) + \frac{e v_{\mathrm{F}}  N_{\mathrm{F}}^{-}}{i\omega}\left[ \mathbf{e}_{z} \times \mathbf{E}(\mathbf{r},\omega) \right]. \label{linresponse}
\end{gathered}
\end{equation}
Where $N_{\mathrm{F}}^{\pm}=N_{\mathrm{F}}^{\uparrow} \pm N_{\mathrm{F}}^{\downarrow}=(p_{\mathrm{F}}^{\uparrow}\pm p_{\mathrm{F}}^{\downarrow})/2\pi\hbar^{2}v_{\mathrm{F}}$ is the sum/difference of the Rashba subband's electronic density of states at the Fermi level and $v_{\mathrm{F}}$ is the Fermi velocity \footnote{We note that both Rashba subbands have the same Fermi velocity for positive Fermi energies \cite{wangPlasmonSpectrumTwodimensional2005}, which is the relevant regime considered in this work.}. It should be noted that these expressions naturally capture the case with only one of two Rashba bands filled --- one of $N_\mathrm{F}^{\uparrow/\downarrow}$ just vanishes. In the absence of spin--orbit interactions, i.e., when two subbands match and  $N_{\mathrm{F}}^{-}=0$, all cross-correlated response functions vanish, such that plasmon and magnon oscillations decouple.

\begin{figure}[h]
    \centering
\includegraphics[clip,trim={0cm 0cm 0cm 0cm},width=0.9\columnwidth]{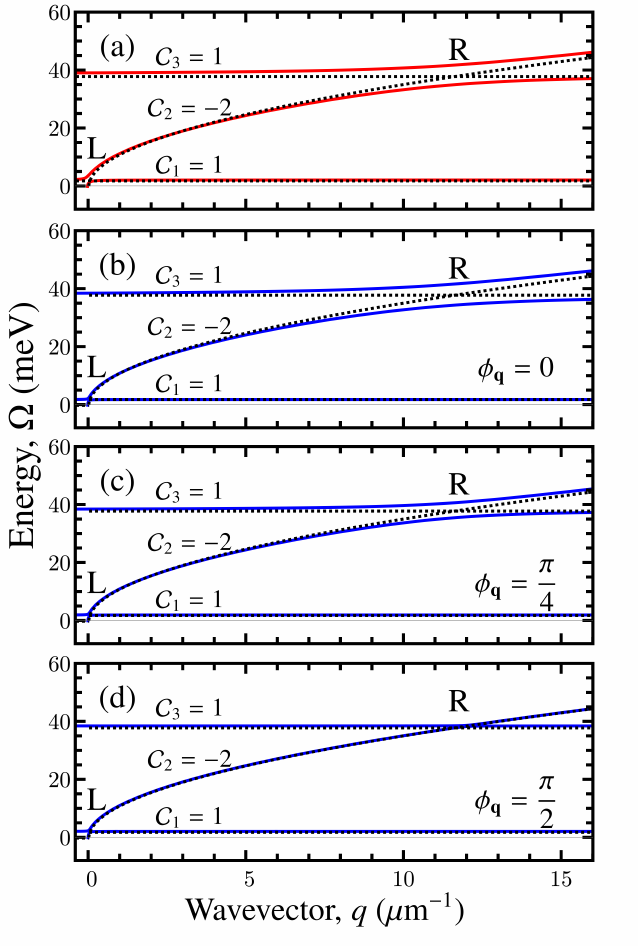}
\vspace{-0.15in}
    \caption{Dispersion of hybrid plasmon--magnon modes (red or blue), shown together with the corresponding bare plasmon and magnon branches (dashed). (a) For \emph{out-of-plane magnetic ordering}, the hybrid modes acquire a nontrivial topological character, with each branch carrying a nonzero Chern number (see insets). For \emph{in-plane magnetic ordering}, the hybridization depends on the relative angle between the propagation direction and the magnetization axis, $\phi_{\mathbf{q}}$, and vanishes for perpendicular orientation, $\phi_{\mathbf{q}}=\pi/2$, leading to Dirac-like branch crossings. In all panels, $\mathrm{L}$ ($\mathrm{R}$) denotes left- (right-) handed magnon modes.}
    \label{Figure_2}
    \vspace{-0.2in}
\end{figure}

The conventional and shortest approach to calculate the dispersion of hybrid modes is to exclude all the fields except for the electric field, yielding a single equation in terms of the dielectric function $\epsilon(\mathbf{q},\omega)=0$ that incorporates electron-magnon interactions. For a realistic set of parameters given in Sec.~\ref{SecIII}, the dispersion of hybrid modes is presented in Fig.~(\ref{Figure_2}a). Plasmons interact effectively with both left- and right-handed magnon modes, spanning the GHz and THz frequency ranges. As is evident, the coupling to the optical mode is stronger than the acoustic one. Although the conventional approach successfully describes the dispersion relations of the hybrid modes, it overlooks their nontrivial topological character.

The nontrivial topological features become apparent only when the set of equations, Eqs~(\ref{LLG}-\ref{linresponse}), are reformulated in terms of an effective Hamiltonian, following the framework developed to describe plasmons in 2D and 3D electron gases \cite{PhysRevB.110.L041406,PhysRevB.111.165404,PhysRevB.105.205426,TopologicalMagnetoplasmon2016}, as well as other physical systems \cite{Perez_Leclerc_Laibe_Delplace_2025,PhysRevResearch.6.033161,doi:10.1126/science.aan8819, PhysRevA.107.023319}. If we exclude the spin density and electric fields and rescale the field describing magnetic fluctuations, $\boldsymbol{\zeta}^{\alpha}(\mathbf{q},\omega) = \sqrt{g_{\alpha}/\bar\omega^{\alpha}_\mathbf{q}} e v_{F} \Delta_{\alpha} N_{F}^{-} \boldsymbol{\ell}^{\alpha}(\mathbf{q},\omega)$, the corresponding equations can be presented as an eigenvalue problem 
$\Omega \psi(\vec{q},\omega)=\hat{\mathcal{H}}_{\mathrm{p-m}}(\vec{q}) \psi(\vec{q},\omega)$.  Here, the $5$-component vector $\psi(\vec{q},\omega)=\{\zeta^{\alpha}_x(\vec{q},\omega),\zeta^{\alpha}_y(\vec{q},\omega), j_x(\vec{q},\omega), \omega_\mathrm{p}(\vec{q}) \rho(\vec{q},\omega)/q, j_y(\vec{q},\omega) \}$ serves as the state vector, and the Hermitian matrix  $\hat{\mathcal{H}}_{\mathrm{p-m}}$
can be interpreted as the ``effective Hamiltonian" of the considered system and is given by  
\begin{equation}
\mathcal{H}_{\vec{q}}^\mathrm{out}=\left(\begin{array}{ccccc}
0 & -i \bar{\omega}^\alpha_{\mathbf{q}} & 2 i M^{\alpha}_{\mathbf{q}} & 0 & 0 \\
i \bar{\omega}^\alpha_{\mathbf{q}} & 0 & 0 & 0 & 2 i M^{\alpha}_{\mathbf{q}} \\
- 2i  M^{\alpha}_{\mathbf{q}} & 0 & 0 & \omega^\mathrm{p}_{\mathbf{q}}n^x_\vec{q} & -i g_{\alpha} \\
0 & 0 & \omega^\mathrm{p}_{\mathbf{q}} n^x_\vec{q} & 0 & \omega^\mathrm{p}_{\mathbf{q}} n^y_\vec{q}\\
0 & - 2 i M^{\alpha}_{\mathbf{q}} & i g_{\alpha} & \omega^\mathrm{p}_{\mathbf{q}} n^y_\vec{q} & 0
\end{array}\right) . \label{EffHam1}
\end{equation}
The bottom-right $3\times 3$ block describes plasmons in the absence of magnons. Here,
$\omega^\mathrm{p}_{\mathbf{q}}=v_{\mathrm{F}}\sqrt{\alpha_{\mathrm{eff}} k_{\mathrm{F}} q}$
is the bare plasmon dispersion with $\alpha_\mathrm{eff}= \alpha_{\mathrm{Vac}}c/\kappa v_{\mathrm{F}}$ being the modified fine-structure constant and $\alpha_{\mathrm{Vac}}$ being the vacuum fine-structure constant. $n_{\vec{q}}=\{\cos (\phi_\vec{q}), \sin (\phi_\vec{q})\}$ denotes its direction of plasmon propagation.
The frequency
$g_{\alpha}=\alpha_{\mathrm{R}}^{2}\Delta_{\alpha}^{2}N_{\mathrm{F}}^{+}/v_{\mathrm{F}}^{2}\rho_{\mathrm{S}}$
quantifies the strength of plasmon coupling to magnetic moments across the interface.
Mimicking the effect of an out-of-plane magnetic field, this coupling opens a gap in the plasmon spectrum,
$\bar{\omega}^\mathrm{p}_{\mathbf{q}}=\sqrt{(\omega^\mathrm{p}_{\mathbf{q}})^2+g_{\alpha}^2}$.
For realistic material parameters, however, this gap remains small and is barely discernible in Fig.~(\ref{Figure_2}). 

The top-right $2\times2$ block describes magnons in the absence of plasmons. Exchange interactions with the electronic spins across the interface renormalize the magnon dispersion,
$\bar{\omega}^\alpha_{\mathbf{q}}=\omega^\alpha_{\mathbf{q}}+g_{\alpha}(v_{\mathrm{F}}^{2}/\alpha_{\mathrm{R}}^{2}-1)$ thereby shifting it to higher frequencies.

The off-diagonal $2\times3$ and $3\times2$ blocks describe the plasmon--magnon coupling. Its matrix structure can be understood as follows: according to the LLG equation, the dynamics of one component of the magnetic fluctuations, i.e. $\ell_x(\vec{r},t)$, is governed by the spin polarization perpendicular to it, $\dot{\ell}_x(\vec{r},t)=\Delta\, s_y(\vec{r},t)/\rho_s$. For a Rashba electronic spectrum, the spin density $s_y(\vec{r},t)$ accompanies the charge current $j_x(\vec{r},t)$ and therefore enforces a component-resolved coupling $\ell_x\leftrightarrow j_x$ (and analogously $\ell_y\leftrightarrow j_y$). The coupling matrix element $M^{\alpha}_{\mathbf{q}}=\sqrt{g_{{\alpha}} \bar\omega^{\alpha}_{\vec{q}}}/2$ depends not only on the coupling frequency $g_{\alpha}$ but also explicitly on the magnon frequency. As a consequence, the plasmon--magnon coupling is stronger for the high-frequency magnon branch than for the low-energy branch, in agreement with Fig.~(\ref{Figure_2}).

The effective Hamiltonian $\mathcal{H}_{\vec{q}}^\mathrm{out}$ contains the complete information about the dispersion of the hybrid plasmon-magnon modes and their topologies. The dispersion relations are given by the positive frequency eigenvalues of the effective Hamiltonian~\footnote{Due to the classical nature of the considered problem --- spin- and current-densities are real fields --- the effective Hamiltonian $\mathcal{H}_{\vec{q}}^\mathrm{out}$ respects the unbreakable particle-hole symmetry. As a result, its negative frequency eigenmodes are not dynamically independent from the positive frequency ones and do not need to be considered separately. Furthermore, $\mathcal{H}_{\vec{q}}^\mathrm{out}$ has a spurious zero-frequency mode protected by the interplay between particle-hole and inversion symmetries of the effective Hamiltonian} and are given by 
\begin{equation}
    \Omega_{\mathbf{q}}^{\pm} = \frac{\tilde{\omega}_{\mathbf{q}}^{\mathrm{p}}+\bar{\omega}_{\mathbf{q}}^{\alpha}}{2} \pm \sqrt{\left(\frac{\tilde{\omega}_{\mathbf{q}}^{\mathrm{p}}-\bar{\omega}_{\mathbf{q}}^{\alpha}}{2}\right)^{2}+(M^{\alpha}_{\mathbf{q}})^{2}}, \label{Dispersion1}
\end{equation}
 and is presented in Fig.~(\ref{Figure_2})~\footnote{We note that Fig.~(\ref{Figure_2}) depicts the hybridization of plasmons with two distinct magnon branches. Although Eq.~(\ref{EffHam1}) includes only a single magnon branch, extending the effective Hamiltonian to incorporate both branches is straightforward.}. The three hybrid branches can be labeled by index $n=1,2,3$ and their topology is characterized by the Chern numbers
\begin{equation}
    \mathcal{C}_n= 
\int\frac{d\mathbf{q}}{2\pi}  \, \mathcal{B}_n(\mathbf{q}), \quad  \mathcal{B}_n(\mathbf{q})=\nabla_{\mathbf{q}} \times\mathcal{A}_n(\vec{q})
\end{equation}
Here we introduce the Berry curvature $\mathcal{B}_n(\mathbf{q})$, defined via the Berry connection
$\mathcal{A}_n(\mathbf{q})=\bra{\mathbf{u}_{\mathbf{q}}^{n}}\, i\nabla_{\mathbf{q}} \,\ket{\mathbf{u}_{\mathbf{q}}^{n}}$, where $\ket{\mathbf{u}_{\mathbf{q}}^{n}}$ denotes the corresponding eigenvector. As shown in Fig.~(\ref{Figure_2}), the set of Chern numbers $\mathcal{C}=\{1,-2,1\}$ is nonzero, reflecting nontrivial topologies of the hybrid modes. We defer the discussion of possible manifestations of these topologies to Sec.~\ref{SecIII} and instead focus on how the nontrivial topologies emerge. 

The interaction-induced mixing between plasmons and magnons is most pronounced in the vicinity of the avoided crossings, where it can be accurately described by effective $2\times 2$ Hamiltonians obtained by projecting the full Hamiltonian onto a pair of the interacting modes. Its explicit expression is \footnote{Really, there exists anomalous couplings between the magnon and plasmon modes in addition to the direct couplings presented in the main text. These couplings can be fully elucidated by projecting the effective Hamiltonian Eq.~(\ref{EffHam1}) onto both the positive and negative frequency eigenmodes. Doing so yields a 4x4 effective Hamiltonian of which has the structure of a Bogoliubov-de-Gennes dynamical matrix:
        \begin{equation}
        \mathcal{H}_{\mathbf{q}} =
        \begin{pmatrix}
            \bar{\omega}_{\mathbf{q}}^{\mathrm{p}} & M_{\mathbf{q}}e^{i\phi_{\mathbf{q}}} & 0 & M_{\mathbf{q}}e^{-i\phi_{\mathbf{q}}} \\
            M_{\mathbf{q}}e^{-i\phi_{\mathbf{q}}} & \bar{\omega}_{\mathbf{q}}^{\mathrm{m}} & -M_{\mathbf{q}}e^{i\phi_{\mathbf{q}}} & 0 \\
            0 & -M_{\mathbf{q}}e^{-i\phi_{\mathbf{q}}} & -\bar{\omega}_{\mathbf{q}}^{\mathrm{p}} & -M_{\mathbf{q}}e^{-i\phi_{\mathbf{q}}} \\
            M_{\mathbf{q}}e^{i\phi_{\mathbf{q}}} & 0 & -M_{\mathbf{q}}e^{i\phi_{\mathbf{q}}} & -\bar{\omega}_{\mathbf{q}}^{\mathrm{m}}
        \end{pmatrix}. \label{AnomCouplingsHamPaper}
    \end{equation}
        The off-diagonal blocks correspond to the anomalous couplings. The anomalous couplings in principle re-normalise the hybrid plasmon-magnon spectrum, but at the exchange coupling strengths considered here the renormalisation is negligible ($\sim\mu \mathrm{eV}$ scales) and thus we can safely discard them.}
\begin{equation}
\bar{\mathcal{H}}_{\vec{q}}^\mathrm{out} = \begin{pmatrix}
\bar{\omega}^\mathrm{p}_{\mathbf{q}} & M^{\alpha}_\vec{q}e^{\alpha i \phi_\vec{q}} \\ M^{\alpha}_\vec{q}e^{-\alpha i \phi_\vec{q}} & \bar{\omega}^{\alpha}_{\mathbf{q}}
    \end{pmatrix}. \label{ReducedEffHam}
\end{equation}
confirms the interpretation of $M^{\alpha}_{\mathbf{q}}$ as the magnitude of the matrix element governing plasmon--magnon coupling. The Hamiltonian $\bar{\mathcal{H}}_{\vec{q}}^\mathrm{out}$ constitutes a canonical minimal model of a band-inversion--driven topological transition, featuring handedness-dependent phase winding factors. The intertwining of plasmons with right- and left-handed modes endows the resulting hybrid bands with Chern numbers $\mathcal C_\mathrm{R} = \{1,-1,0\}$ and $\mathcal C_\mathrm{L} = \{0,-1,1\}$, respectively. Taken together, these contributions reproduce the result given for $\mathcal{C}$ presented above.

Typically, a $2\times 2$ Hamiltonian associated with $\bar{\mathcal{H}}_{\vec{q}}^{\mathrm{out}}$ describes the hybridization between an $s$-wave-like band and an opposite-parity band of $p_{x}\pm i p_y$ symmetry. Such a structure is characteristic of band inversion problems underlying topological phase transitions. In the present setting, the role of the chiral $p_x\pm i p_y$ branch is played by magnons: the circular precession of localized spins endows them with a helicity-dependent orbital character. The longitudinal plasmon mode, by contrast, is even under parity and therefore serves as the effective $s$-wave sector of the hybridized system.
\subsection{In-plane magnetic ordering}
\label{SecIIb}
The framework developed above can be straightforwardly extended to ferrimagnets with in-plane anisotropy. Without loss of generality, we take the equilibrium spin-ordering direction, represented by the unit vector $\vec{n}$, to lie along the $y$ axis. Magnetization fluctuations, described by $\boldsymbol{\ell} = (\ell_x, 0, \ell_z)$, then involve both in-plane and out-of-plane components, whose physical roles are not equivalent. 

Unless the Fermi energy is extremely small --- which is unfavorable for plasmons --- the out-of-plane component of the electronic spin near the Fermi level is not only weak in magnitude but also aligned in the same direction across the entire Fermi contour in both bands. As a result, uniform plasmon-induced current oscillations do not produce an out-of-plane spin-density response, and the coupling to the $\ell_z$ component of magnetic fluctuations is therefore suppressed in the long-wavelength limit. The plasmon-magnon interaction is thus mediated only via the linear response of the in-plane component of the magnetic fluctuations $\ell_x$ to the corresponding electronic spin-density component $s_x$.

The extension of the effective Hamiltonian framework for in-plane magnetic ordering of ferrimagnets is presented in Appendix~\ref{App.B}. It can be characterized by the effective Hamiltonian acting at the $5$-component vector $\psi(\vec{q},\omega)=\{\zeta_x(\vec{q},\omega),\zeta_z(\vec{q},\omega), j_x
(\vec{q},\omega), \omega_\mathrm{p}(\vec{q}) \rho(\vec{q},\omega)/q, j_y(\vec{q},\omega) \}$
 and given by 
\begin{equation}
\mathcal{H}_{\vec{q}}^\mathrm{in}=\left(\begin{array}{ccccc}
0 & -i \bar{\omega}^\alpha_{\mathbf{q}} & 0 & 0 & 0 \\
i \bar{\omega}^\alpha_{\mathbf{q}} & 0 & 0 & 0 & 2i M^{\alpha}_{\mathbf{q}} \\
0 & 0 & 0 & \omega^\mathrm{p}_{\mathbf{q}} n^x_\vec{q} & 0\\
0 & 0 & \omega^\mathrm{p}_{\mathbf{q}} n^x_\vec{q} & 0 & \omega^\mathrm{p}_{\mathbf{q}}n^y_\vec{q} \\
0 & -2i M^{\alpha}_{\mathbf{q}} & 0 & \omega^\mathrm{p}_{\mathbf{q}} n^y_\vec{q} & 0
\end{array}\right) . \label{EffHam2}
\end{equation}
It has the same mathematical structure as $\mathcal{H}_{\vec{q}}^\mathrm{out}$, but only the coupling $\ell_z\leftrightarrow j_y$ is involved, and the dispersion relation of plasmons is not renormalized by interactions with spins across the interface.
We use the same set of parameters for plotting, and the resulting dispersion relations of the hybrid modes are shown in Fig.~\ref{Figure_2}(b-d). In contrast to the case of out-of-plane magnetic ordering, the spectra are strongly anisotropic and exhibit the same angular dependence for both magnon branches. The plasmon-magnon coupling is maximized when the plasmon propagates parallel to the magnetic ordering, and vanishes for perpendicular propagation, where the plasmon-induced spin-density modulation is orthogonal to the magnetic fluctuations. This angular dependence can be made explicit by projecting the full Hamiltonian onto the pair of interacting hybrid modes, yielding
\begin{equation}
\bar{\mathcal{H}}_{\vec{q}}^\mathrm{in} = \begin{pmatrix}
\bar{\omega}^\mathrm{p}_{\mathbf{q}} &  M^{\alpha}_{\mathbf{q}}\cos(\phi_\vec{q}) \\
M^{\alpha}_{\mathbf{q}}\cos(\phi_\vec{q}) & \bar{\omega}^\alpha_{\mathbf{q}} \end{pmatrix}.
\end{equation}
The angular dependence of the matrix element \(M^{\alpha}_{\vec q}\cos\phi_{\vec q}\) exhibits \(p\)-wave symmetry and vanishes at \(\phi_{\vec q}=\pm \pi/2\). In the vicinity of the branch-touching points, the effective Hamiltonian can be linearized as \(\bar{\mathcal H}^{\mathrm{out}}_{\vec q}=\epsilon_0 \pm v_x(\hat q_x-q_0)(\hat 1+\hat\sigma_z)/2+ v_y \hat q_y\), revealing the Dirac-like nature of the branch crossings. Although the emergent Dirac points are not protected by an exact symmetry, the branch crossings are exact within the leading-order linearized theory considered here. Within this reduced polarization sector, the existence of the nodes is therefore a robust consequence of the polarization structure and is independent of the specific approximations used to evaluate the electronic response functions. The direct electromagnetic coupling mediated by the oscillating magnetic field cannot gap the nodes either: besides being small in magnitude, it has the same \(p\)-wave symmetry angular dependence as the exchange-mediated coupling and consequently vanishes along the same nodal directions (see App.~\ref{DirectCoupling} for a detailed discussion). A gap can possibly arise only from subleading effects beyond the present treatment. In particular, the out-of-plane component of the magnon polarization is decoupled at leading order in $v_{\mathrm{F}}q/\omega$, because the plasmon carries no macroscopic out-of-plane spin polarization. The corresponding coupling may, however, reappear at second and higher orders in $v_{\mathrm{F}}q/\omega$. For the experimental parameters considered here and within the frequency-momentum range near the plasmon-magnon avoided crossings, $(v_{\mathrm{F}}q/\omega)^2\approx0.04$. Any resulting gap at the Dirac nodes is therefore expected to be very small.
\section{Discussions}
\label{SecIII}
Owing to the hybrid nature of the modes, the avoided crossing can be accessed through either their plasmonic or magnonic components. In particular, the plasmonic sector can be probed using scattering-type near-field optical microscopy (s-SNOM), as well as other related near-field techniques \cite{feiGatetuningGraphenePlasmons2012,doi:10.1126/sciadv.adu1415, grigorenkoGraphenePlasmonics2012a}, or through electron energy-loss spectroscopy \cite{PhysRevB.77.233406,PhysRevB.82.201413}. Conversely, probing the magnonic sector --- for example via inelastic neutron scattering --- will reveal the same avoided-crossing structure in the magnon spectrum, reflecting the underlying hybridization of the modes \cite{PhysRevB.101.054403,ItohShinichi2016Wfas}.

The experimental observation of the predicted hybrid plasmon--magnon modes requires the linewidth of the hybrid modes to remain sufficiently small for the hybridization gap to be clearly resolved. This linewidth receives contributions from both the plasmonic and magnonic sectors. In high-quality ferrimagnetic materials such as the YIG considered here, the magnon linewidth is exceptionally narrow. Ferrimagnetic resonance linewidths of the order of $\sim 1~\mathrm{MHz}$ have been reported \cite{10.1063/1.5115266}, corresponding to an energy broadening of only a few $\mathrm{neV}$. This is approximately five to six orders of magnitude smaller than the predicted hybridization gap, rendering the magnonic contribution to the linewidth negligible. The dominant broadening therefore originates from the plasmonic sector, where even in clean, encapsulated graphene the characteristic plasmon linewidth is of the order of $1$--$2~\mathrm{meV}$~\cite{doSlowHighlyConfined2025,dajornadaUniversalSlowPlasmons2020, woessnerHighlyConfinedLowloss2015a}. This linewidth is determined by impurity and defect scattering, electron--acoustic phonon scattering, and electron--electron interactions. An additional source of broadening is Landau damping. However, in the ferrimagnetic heterostructures considered here, the avoided crossing occurs sufficiently far from the particle--hole continuum that thermal broadening at finite temperatures does not lead to appreciable Landau damping. Recent theoretical and experimental advances have established Janus transition-metal dichalcogenides as a high-quality platform for two-dimensional electronics and optoelectronics \cite{https://doi.org/10.1002/pssb.202100562,luJanusMonolayersTransition2017}, while highly confined, low-loss plasmons have recently been demonstrated in closely related atomically thin TMD systems~\cite{doSlowHighlyConfined2025}. These developments suggest that electrostatically doped Janus TMDs provide a promising platform for realising the hybrid plasmon--magnon modes predicted here.

To probe the anisotropic hybridisation gap of the in-plane easy axis configuration, we expect a scattering-type near field optical setup with a mobile tip \cite{chenOpticalNanoimagingGatetunable2012, feiGatetuningGraphenePlasmons2012} can probe the directional dependence of the hybridisation gap. The mobile tip can be scanned across the sample, exciting plasmons with different propagation directions, thus when observing the plasmonic excitation spectrum, an anisotropic gap should be observed, with different sizes depending on the propagation direction of the excited plasmon.

It is instructive to summarize the principal approximations underlying our theory and their range of validity. First, the electronic response is treated within the random phase approximation (RPA), the standard theoretical framework for describing collective charge excitations in doped two-dimensional electron systems. Beyond-RPA exchange and correlation effects are expected to produce only quantitative renormalizations of the plasmon dispersion and damping. Second, retardation effects are neglected, as the hybrid modes lie well outside the light cone where such effects become important. Finally, the magnetization dynamics is described by the Landau--Lifshitz--Gilbert equation, which provides an excellent description of low-energy spin dynamics in ferrimagnetic insulators such as YIG. Quantum corrections are suppressed in the large-spin limit and are expected to result only in quantitative renormalizations of the magnon spectrum. Finally, we neglect the direct relativistic coupling between plasmons and magnons mediated by the oscillating magnetic field of the plasmon. For the systems considered here, this interaction is expected to be much weaker than the interfacial exchange mechanism and therefore contributes only small quantitative corrections to the hybridization (see App.~\ref{DirectCoupling} for estimations).

A hallmark of nontrivial topology is the existence of robust, topologically protected edge modes at interfaces between regions characterized by different Chern numbers. Reversing the direction of the equilibrium out-of-plane magnetization ($\mathbf{e}_z \rightarrow -\mathbf{e}_z$) in a ferrimagnet reverses the precession of the magnetic moments and changes the sign of the Chern numbers for all three hybrid branches ($\mathcal{C} \rightarrow -\mathcal{C}$). The existence of truly protected interface states, however, also requires global bulk gaps in the spectra on both sides of the interface. In the absence of such gaps, the bulk--boundary correspondence neither guarantees the existence of trapped modes at magnetic domain walls nor, if such modes exist, ensures their topological protection. This is the case for an isolated domain wall separating regions with uniform out-of-plane magnetization. A global gap can, however, be engineered through periodic magnetic patterning, for example in magnon crystals~\cite{PhysRevLett.109.137202,gubbiottiSpinWaveBand2013}. This situation is analogous to other hybrid systems, such as exciton--polaritons, where the topological properties become fully manifest only after introducing a suitable periodic structure~\cite{PhysRevX.5.031001}.

The emergence of topological magnon-plasmon modes is accompanied by a sizable Berry curvature across all three hybrid branches. The Berry curvature is strongly concentrated near the avoided crossings and can give rise to a pronounced anomalous thermal Hall response~\cite{ZHANG20241}. Similar effects have been discussed for hybrid phonon-magnon systems \cite{MagnonPhononNew1,MagnonPhononNew2}. Experimental observation of such a thermal Hall signal would therefore provide a clear signature of the formation of topological hybrid modes.

For material estimates, the magnonic sector parameters are relevant to the YiG ferrimagnetic insulator \cite{princepFullMagnonSpectrum2017}: We approximate the bare magnon dispersion to be flat and given by $\bar\omega_{\mathbf{q}}^{\mathrm{R}(\mathrm{L})}\approx 38\ \mathrm{meV} \ (2 \ \mathrm{meV})$, a density of magnetic spins of $\rho_{\mathrm{S}} \approx 2 \times 10^{12} \ \mathrm{cm}^{-2}$, and an effective exchange interaction of $\Delta_{\mathrm{R}(\mathrm{L})}=70 \ \mathrm{meV} \ (50\ \mathrm{meV})$ (See App.~\ref{MICMagnons} for details). The electron sector parameters are relevant to the Janus two-dimensional dichalcogenide SMoSe \cite{zhangJanusMonolayerTransitionMetal2017}: electron doping of $n_{e} \approx 2.5 \times 10^{12} \ \mathrm{cm}^{-2}$, a Rashba coupling strength of $\alpha_{\mathrm{R}} = 1.5 \times 10^{-5} \ \mathrm{meV} \ \mathrm{cm}$, an effective electron mass of $m = 0.05 m_{0}$, where $m_{0}$ is the electron rest mass, and a relative dielectric constant of $\kappa = 3$.

Employing ferrimagnets in magnetic heterostructures enhances the magnon-plasmon coupling strength, which can be seen by calculating the coupling matrix elements. With parameters stated above, the resulting matrix elements are given by $M_{\mathbf{q}}^{\mathrm{R}} \approx 3.64 ~\mathrm{meV}$ and $M_{\mathbf{q}}^{\mathrm{L}} \approx 0.6~\mathrm{meV}$, showing that the coupling with the high-frequency $\mathrm{THz}$ magnon mode is strongly enhanced compared to that of the lower frequency mode, which is evident in Fig.~(\ref{Figure_2}). This enhancement is due to a combination of the magnon energy dependence ($\propto\sqrt{\bar{\omega}_{\mathbf{q}}^{\alpha}}$) of the matrix elements, which favors the higher frequency magnon mode, and the s-d(f) onsite exchange interaction between electron and magnetic spin becoming enhanced by a factor dependent on the Bogoliubov coefficients (see App.~\ref{MICMagnons} for the microscopic derivation). The latter is a feature that ferrimagnets share with antiferromagnets \cite{dyrdalMagnonplasmonHybridizationMediated2023, Shiranzaei_2022}, but is distinguished from antiferromagnets by an inherent broken degeneracy between the magnon chiralities that pushes the upper magnon mode into the $\mathrm{THz}$ frequency range, which is favorable for plasmonic experiments. The enhanced coupling also enables access to the strong-hybridisation regime in which the hybridization gap exceeds plasmonic spectral broadening induced by dissipative processes such as Landau damping and impurity scattering~\cite{FeiZ2012Gogp, chenOpticalNanoimagingGatetunable2012a}. Thus, ferrimagnet-based magnetic heterostructures provide a readily accessible experimental platform for investigating plasmon-magnon hybridization, which has not yet been experimentally demonstrated.

A unique feature of the ferrimagnet employed in the heterostructure considered here is that both the optical and acoustic magnon modes carry a finite net spin polarization, allowing them to couple efficiently to plasmons via the interfacial exchange mechanism considered in this work. In contrast, in multisublattice ferromagnets and antiferromagnets, the optical magnon modes generally carry zero net spin polarization in the absence of an external magnetic field. Consequently, the exchange mechanism considered here does not give rise to plasmon--magnon coupling for these modes. A residual coupling may still arise through the direct relativistic interaction mediated by the oscillating magnetic field accompanying the plasmon, although the resulting hybridization is expected to be significantly weaker (see App.~\ref{DirectCoupling} for details).

To conclude, by combining microscopic and macroscopic descriptions of magnon-plasmon coupling, we demonstrate that plasmons and magnons undergo strong hybridization at the interface between a Janus TMD and an insulating two-sublattice ferrimagnet, spanning both the GHz and THz frequency regimes. Moreover, the developed framework provides a natural and efficient classification of the topology of the hybrid modes, which originates from the phase winding of the magnon-plasmon coupling induced by spin-momentum locking and the accompanying chiral winding of the electronic spin along the Fermi contours.

\textit{Note added:} After submission of our work, we became aware of related studies on topological plasmon-magnon modes in magnetic heterostructures involving ferromagnets, antiferromagnets, and skyrmion crystals \cite{hirosawa2025topologicalmagnonplasmonhybrids, gunnink2026couplingplasmonstwomagnoncontinuum}. While the computational approaches differ, the conclusions are closely aligned. 

\section*{Acknowledgements}
We acknowledge fruitful discussions with Michael Fuhrer, Haoran Ren, Victor Gurarie and Michail Glazov, as well as support from the Australian Research Council Centre of Excellence in Future Low-Energy Electronics Technologies (CE170100039).

\appendix
\section{Microscopically evaluated response functions} \label{Ap.1}
This Appendix presents a microscopic derivation of the expression Eq.~(\ref{linresponse}) from the main text.  The Rashba electrons can be described by the following Hamiltonian
\begin{equation}
    \mathcal{H}_{0} = \frac{\mathbf{p}^{2}}{2m} + \alpha_{\mathrm{R}}\left[ \mathbf{p} \times \boldsymbol{\sigma}\right]_{z} +\Delta \sigma_{z}, \label{RashbaHamiltonian}
\end{equation}
Here $m_{*}$ is the effective electron band mass, $\alpha_{\mathrm{R}}$ is the Rashba coupling, and $\Delta$ is the exchange energy from the coupling with the Ferrimagnet. We note that the magnetic exchange interaction term in Eq.~(\ref{RashbaHamiltonian}) implies the magnetisation axis to be pointing in the z-direction. This does not lose any generality with the calculation of the response functions, and it is easy to generalise the following calculations to the in-plane magnetisation axis case. The electron dispersion from this Hamiltonian is given as $\epsilon_{\mathbf{p}}^{\sigma} = \frac{\hat{\mathbf{p}}}{2m_{*}} + \sigma \epsilon_{\mathbf{p}}^{\alpha_{\mathrm{R}}}$, $\epsilon_{\mathbf{p}}^{\alpha_{\mathrm{R}}}=  \sqrt{\alpha_{\mathrm{R}}^{2}  p^{2}+\Delta^{2}}$ is the spin-orbit interaction energy, and $\sigma = \uparrow/\downarrow$ indexes the Rashba subbands with opposite spin polarisation. The spinor wavefunctions of this Hamiltonian are given as
\begin{equation}
|\uparrow, \mathbf{p} \rangle=\binom{\cos \left(\frac{\theta}{2}\right)}{i \sin \left(\frac{\theta}{2}\right) e^{i \phi_{ \mathbf{p}}}}, \quad|\downarrow, \mathbf{p} \rangle=\binom{\sin \left(\frac{\theta}{2}\right)}{-i \cos \left(\frac{\theta}{2}\right) e^{i \phi_{ \mathbf{p} }}} .
\end{equation}
Here $\phi_{\mathbf{p}}$ is the polar angle for the momentum $\mathbf{p}$ and $\cos(\theta) = \Delta/\epsilon_{\mathbf{p}}^{\alpha_{\mathrm{R}}}$ is the out-of-plane spin component.\\
The calculation of the susceptibilities is essentially the calculation of the corresponding spin-current polarisation bubbles. First we note that the spin and current operators of the Hamiltonian Eq.~(\ref{RashbaHamiltonian}) are given by $\hat{j} = \frac{e}{m_{*}}\mathbf{\hat{p}}+ e\alpha_{\mathrm{R}}\left[\boldsymbol{\sigma} \times \mathbf{e}_{z} \right]$, and $\hat{s} = \left[\boldsymbol{\sigma} \times \mathbf{e}_{z} \right]$. We thus can express the linear response of the current and density fields as the following
\begin{equation}
    \begin{gathered}
        \mathbf{j}(\mathbf{r},\omega) = \hat{\sigma}(\omega) \mathbf{E}(\mathbf{r},\omega) + \hat{\chi}(\omega)\left[\mathbf{e}_{z} \times \boldsymbol{\ell}(\mathbf{r},\omega) \right], \\
        \mathbf{s}(\mathbf{r},\omega) = \hat{\kappa}(\omega)\boldsymbol{\ell}(\mathbf{r},\omega)+\hat{\Lambda}(\omega)\left[\mathbf{e}_{z} \times \mathbf{E}(\mathbf{r},\omega) \right]. \label{ResponseFunc}
    \end{gathered}
\end{equation}
\begin{figure}[h]
    \centering
    \includegraphics[width=1\linewidth]{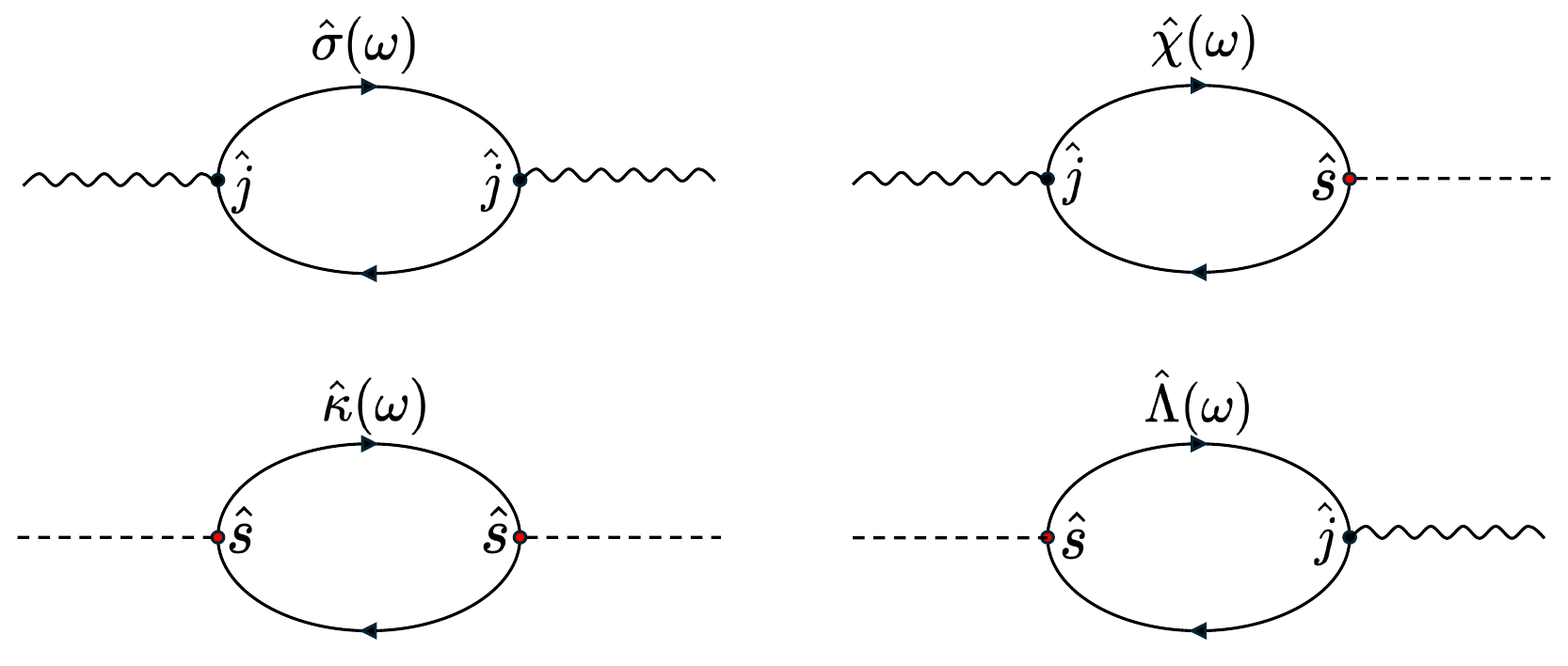}
    \caption{Feynmann diagrams of the response functions in Eq.~(\ref{ResponseFunc}). The solid arrowed lines represent the non-interacting retarded electron Greens function $G^{(0)}_{\sigma}(\mathbf{q},\omega)=(\omega - \epsilon_{\mathbf{p}}^{\sigma}+i\delta)^{-1}$, the dashed line represents the complex free magnon propagator $\mathcal{D}^{(0)}(\mathbf{q},\omega)=-2\varepsilon_{\mathbf{q}}/((\omega+i\delta)^{2}+\varepsilon_{\mathbf{q}}^{2})$, and the wiggly line represents the free plasmon field. Counter-clockwise from the top left: $\hat{\sigma}(\omega)$ is the AC conductivity given by the electron current-current correlation function. $\hat{\chi}(\omega)$ is the response function which encodes the current response from the magnetic spin fluctuation, given by the electron current-spin correlation function. $\hat{\kappa}(\omega)$ is the response function which encodes the electron spin-density response to the magnetic spin fluctuation, given by the spin-spin correlation function. $\hat{\Lambda}(\omega)$ is the response function which encodes the spin-density response to the internal self consistent electric field from the electron density fluctuations, given by the spin-current correlation function.}
    \label{APP.A Fig 1}
\end{figure}
Fig.~(\ref{APP.A Fig 1}) shows the relevant Feynmann diagrams needed to calculate the response functions. Starting with the AC conductivity, we note that as there are no external magnetic fields, the conductivity tensor is diagonal with only the longitudinal conductivity being non-zero, $\hat{\sigma}(\omega)=\sigma_{xx}(\omega)$. The AC conductivity is directly proportional to the density-density response function and thus can be calculated in the RPA using the formalism developed in \cite{PhysRevB.75.205418,wunschDynamicalPolarizationGraphene2006,stauberPlasmonicsDiracSystems2014a} as
\begin{equation}
    \begin{aligned}
     \sigma_{xx}=\frac{i}{\omega}\sum_{\mathbf{p},\sigma} |\bra{\mathbf{p+q},\sigma}\hat{j}_{x}\ket{\mathbf{p},\sigma}|^{2} \mathcal{Z}
    \end{aligned}
\end{equation}
Here, we have introduced a compact notation
\begin{equation}
\mathcal{Z}=\frac{n_{\mathrm{F}}(\epsilon^{\sigma}_{\mathbf{p+q}})-n_{\mathrm{F}}(\epsilon_{\mathbf{p}}^{\sigma})}{\omega +\epsilon_{\mathbf{p+q}}^{\sigma}-\epsilon_{\mathbf{p}}^{\sigma}+i\delta}\end{equation} Furthermore we have neglected interband particle-hole excitations as for small momenta they become negligibly small \cite{wangPlasmonSpectrumTwodimensional2005}. Henceforth, all calculations will be performed in the limit $q \ll p_{\mathrm{F}}^{\sigma}$ and $\omega \ll \epsilon_{\mathrm{F}}$. This allows the following simplifications
\begin{equation}
\begin{gathered}
    n_F\left(\epsilon_{ \mathbf{p + q} }^\sigma\right)-n_F\left(\epsilon_{ \mathbf{p} }^\sigma\right) = q \cos(\phi_{\mathbf{p}}-\phi_{\mathbf{q}}) \delta\left(p-p_{F}^\sigma\right), \\\epsilon_{ \mathbf{p + q} }^\sigma-\epsilon_{ \mathbf{p} }^\sigma= v_{F} q \cos( \phi_{\mathbf{p}}-\phi_{\mathbf{q}}),
\end{gathered}
\end{equation}
where $v_{\mathrm{F}} = \frac{p_{\mathrm{F}}^{\sigma}}{m_{*}} - \sigma \alpha_{R} = \sqrt{2 m_{*} \epsilon_{\mathrm{F}}+k_{\alpha_{R}}^{2}}$ is the Fermi velocity. With, $\phi = \phi_{\mathbf{p}}-\phi_{\mathbf{q}}$, these simplifications give the following form for the matrix elements,
\begin{equation}
    \bra{\mathbf{p+q},\sigma} \hat{j}_{x}\ket{\mathbf{p},\sigma}   = e v_{\mathrm{F}}\cos(\phi).
\end{equation}
The final simplification we make is that $\omega \gg v_{\mathrm{F}}q$, i.e. the dispersion of hybrid modes is far away from the intraband continuum of particle hole excitations. This is valid because the energy at which the magnon and plasmon branches cross is far away from the particle hole continuum, and thus the essential physics is capture in the aforementioned limit. After this simplifcation, and upon performing the momentum integration we are left with the following form of the AC conductivity
\begin{equation}
    \sigma_{xx}(\omega) = -\frac{e^{2}v_{\mathrm{F}}^{2}N_{F}^{+}}{i\omega}
\end{equation}
The calculation of the other susceptibilities proceeds in a near identical manner, with the following definitions of each susceptibility tensor

\begin{equation}
\begin{split}
    \hat{\kappa}(\omega) = \Delta\sum_{\mathbf{p},\sigma}|\bra{\mathbf{p+q},\sigma}\hat{s}_{i}\ket{\mathbf{p},\sigma}|^{2} \mathcal{Z} \\
    \hat{\Lambda}(\omega) =  \frac{i}{\omega}\sum_{\mathbf{p},\sigma} \bra{\mathbf{p+q},\sigma}\hat{s}_{i} \ket{\mathbf{p},\sigma}\bra{\mathbf{p},\sigma}\hat{j}_{j}\ket{\mathbf{p+q},\sigma} \mathcal{Z}, \\ \hat{\chi}(\omega) = \Delta\sum_{\mathbf{p},\sigma} \bra{\mathbf{p+q},\sigma}\hat{j}_{i}\ket{\mathbf{p},\sigma}\bra{\mathbf{p},\sigma} \hat{s}_{j}\ket{\mathbf{p+q},\sigma} \mathcal{Z}
\end{split}
\end{equation}
and
The final forms of all these susceptibility tensors is stated in Eq.~(\ref{linresponse}).
\section{High Doping regime}
This Appendix presents hybrid modes calculated for higher charge carried densities. The couplings have a $\sqrt{q}$ dependence, and thus the strength of hybridisation are doping dependent. The calculations presented in the main text were performed at a moderate doping of $n_{\mathrm{e}} \approx 2.5 \times 10^{12}~ \mathrm{cm}^{-2}$, corresponding to a Fermi energy of $\epsilon_{\mathrm{F}} \approx 120 ~ \mathrm{meV}$. Here, we perform calculations of the hybrid spectrum at a higher doping of $n_{\mathrm{e}} \approx 7 \times 10^{12}~ \mathrm{cm}^{-2}$, with a corresponding Fermi energy of $\epsilon_{\mathrm{F}} \approx 340 ~ \mathrm{meV}$. This reduces the coupling matrix elements to $M_{\mathbf{q}}^{\mathrm{R}} \approx 2.2 ~ \mathrm{meV}$ and $M_{\mathbf{q}}^{\mathrm{L}} \approx 0.5 ~ \mathrm{meV}$ for the right and left handed chiralities of magnons respectively. However, this is a moderate reduction of the coupling as compared to the main text values of $M_{\mathbf{q}}^{\mathrm{R}} \approx 3.64 ~\mathrm{meV}$ and $M_{\mathbf{q}}^{\mathrm{L}} \approx 0.6~\mathrm{meV}$, respectively, and thus only has a moderate affect on the hybrid mode dispersions.
\begin{figure}
    \centering
    \includegraphics[width=\columnwidth]{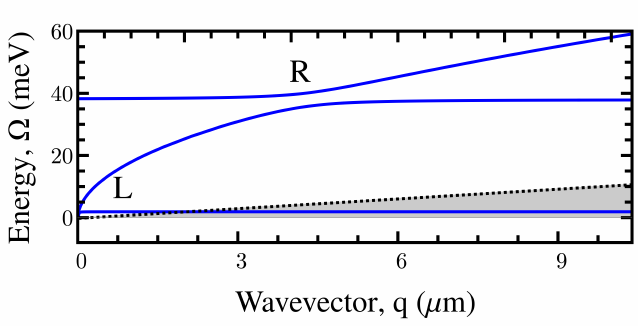}
    \caption{Dispersion of the hybrid spectrum of plasmon-magnon hybrid modes with a charge carrier density of $n_{\mathrm{e}} \approx 7 \times 10^{12}~ \mathrm{cm}^{-2}$, shown with the corresponding interband particle-hole continuum. All other model parameters are the same as in the main text.}
    \label{highdensityfig}
\end{figure}

The hybrid spectrum for this high doping is presented in Fig.~(\ref{highdensityfig}), and it can be seen that in particular the hybridisation between the plasmons and the high right-handed magnon mode remains quite large, and well within the realm of experimental feasibility. Furthermore, we see that the hybrid modes remain well-separated from the particle-hole continuum, and thus avoids significant Landau damping which could obscure the hybridisation gap in spectroscopic probes.

\section{Extension of the framework to the in-plane magnetic ordering}\label{App.B}
This Appendix presents the eigenvalue reformulation of the LLG equations to the in-plane easy axis anisotropy case. Focusing on a single magnon branch, we can express the LLG equation as follows
\begin{equation}
\begin{split}\omega\boldsymbol{\ell}^{\alpha}(\mathbf{r}, t)=-i\omega^{\alpha}_{\boldsymbol{q}}\left[\mathbf{e}_{y}\times\boldsymbol{\ell}^{\alpha}(\mathbf{r}, t)\right]-\\ i\frac{\Delta_{\alpha}}{\rho_{S}} \left[\mathbf{e}_{y}\times\vec{s}(\vec{r},t)\right], \label{LLGHam}
\end{split}
\end{equation}
where, without loss of generality, we chose a magnetization axis of $\mathbf{n}=\mathbf{e}_{y}$. The difference in sign as compared to the out-of-plane LLG equation comes from the fact that we have performed an odd permutation of the basis vectors of the magnitisation vector, so the cross products pick up a negative sign as pseudovectors. Furthermore, the current and spin-density fields have the linear response relation.
\begin{equation}
\begin{gathered}
\mathbf{j} ( \mathbf{r} , \omega)=-\frac{e^2 v_{F}^2 N_{F}^{+}}{i \omega} \mathbf{E} ( \mathbf{r} , \omega)+e v_{F} \Delta_{\alpha} N_{F}^{-}\ell_{x}^{\alpha}(\mathbf{r},t) \mathbf{e}_{y}\\
\mathbf{s} ( \mathbf{r} , \omega)=\Delta_{\alpha} N_{F}^{+}\ell_{x}^{\alpha}(\mathbf{r},\omega) \mathbf{e}_{x} +\frac{e v_{F} N_{F}^{-}}{ i\omega} \left[\mathbf{e}_{z}\times \mathbf{E} ( \mathbf{r} , \omega)\right]. \label{linresponseLLGInPlane}
\end{gathered}
\end{equation}
By excluding the spin density and electric fields in  Eq.~(\ref{LLGHam}) using the linear response relations in Eq.~(\ref{linresponseLLGInPlane}), the LLG equation takes the final form
\begin{equation}
    \omega \boldsymbol{\zeta}^{\alpha}(\mathbf{q},\omega) = -i\bar{\omega}^{\alpha}_{\mathbf{q}}\left[\mathbf{e}_{y}  \times  \boldsymbol{\zeta}^{\alpha}(\mathbf{q},\omega)\right] + 2i M_{\mathbf{q}}^{\alpha}\mathrm{j}_{y}(\mathbf{q},\omega)\mathbf{e}_{z},\label{LLGHamiltonianForm}
\end{equation}  
where we have rescaled the magnetic fluctuation fields to have units of current density as $\boldsymbol{\zeta}^{\alpha}(\mathbf{q},\omega) = \sqrt{g_{\alpha}/\bar\omega^{\alpha}_\mathbf{q}} e v_{F} \Delta_{\alpha} N_{F}^{-} \boldsymbol{\ell}^{\alpha}(\mathbf{q},\omega)$. Finally, we can express the relations for the current and charge density fields by supplementing the expressions with the continuity equation, Eq.~(\ref{Continuity}), which yields the Hamiltonian form of the expressions
\begin{equation}
    \begin{gathered}
        \omega \mathbf{j}(\mathbf{q}, \omega)  =\omega^\mathrm{p}_\vec{q} \hat{\mathbf{n}}_{\mathbf{q}} j_{0}(\mathbf{q},\omega)-2iM_{\mathbf{q}}^{\alpha}\zeta_{z}^{\alpha}(\mathbf{q},\omega)\mathbf{e}_{y} \\
 \omega j_{0}(\mathbf{q},\omega)=\omega^\mathrm{p}_\vec{q} \hat{\mathbf{n}}_{\mathbf{q}} \cdot \mathbf{j}(\mathbf{q}, \omega),
    \end{gathered} \label{ElectronHamForm}
\end{equation}
where $j_{0}(\mathbf{q},\omega)=\omega^\mathrm{p}_\vec{q} \rho(\vec{q},\omega)/q$. Thus, organizing Eqs.~(\ref{LLGHamiltonianForm}) \& (\ref{ElectronHamForm}) into an eigenvalue problem, yields the effective Hamiltonian in Eq.~(\ref{EffHam2})
\section{Emergent Dirac nodes robustness to direct electromagnetic coupling} \label{DirectCoupling}
In this Appendix, we show that the emergent Dirac nodes in the spectrum of hybrid plasmon-magnon modes in the setup involving in-plane easy axis anisotropy are robust to their direct electromagnetic coupling. The latter can be incorporated in the LLG equation as 
\begin{equation}
\begin{gathered}
    \rho_{\mathrm{S}}\left(\alpha \partial_{t}\left[\boldsymbol{\ell}^{\alpha}(\mathbf{r},t) \times \mathbf{e}_{z}\right] - \omega_{\mathbf{q}}^{\alpha} \boldsymbol{\ell}^{\alpha}(\mathbf{r},t) \right)\\
    = \Delta_{\alpha}\mathbf{s}(\mathbf{r},t)+\rho_{\mathrm{S}}\mu_{\alpha}^{\mathrm{B}}\mathbf{B}(\mathbf{r},t),
    \end{gathered}
\end{equation}
where $B(\vec{r},t)$ is the magnetic field created by oscillating electric current density and $\mu_{\alpha}$ is the magnetic moments of oscillating spins. The components of the magnetic field are found by solving Maxwell's equations, and can be presented as 
\begin{equation}
\begin{aligned}
   \mathbf{B}(\mathbf{r},t) = \frac{2\pi}{c} \left[\mathbf{e}_{z}\times \mathbf{j}(\mathbf{r},t) \right]
\end{aligned}
\end{equation}
Following the same computational scheme as done in App.~\ref{App.B}, the in-plane LLG equations show that only the x-component of the magnetic field contributes to the magnetisation dynamics, and they are presented as
\begin{equation}
\begin{gathered}
    \omega ~\ell_{x}^{\alpha}(\mathbf{r},\omega) = -i\omega_{\mathbf{q}}^{\alpha} \ell_{z}^{\alpha}(\mathbf{r},\omega), \\
    \omega \ell_z^\alpha(\mathbf{r},\omega)=i \omega_{ q }^\alpha \ell_x^\alpha(\mathbf{r},\omega)+i \frac{\Delta_\alpha}{\rho_S} s_x(\mathbf{r},\omega)+i \frac{2 \pi\mu_\alpha}{c}   j_y(\mathbf{r},\omega) . \label{Response_Func2}
\end{gathered}
\end{equation}
The current density field, $j_{y}$, can be excluded with the help of Eq.~(\ref{linresponseLLGInPlane}) as
\begin{equation}
\begin{gathered}
    j_{y}(\mathbf{r},\omega)=\frac{e v_{\mathrm{F}}N_{\mathrm{F}}^{+}}{N_{\mathrm{F}}^{-}}s_{x}(\mathbf{r},\omega)+ev_{\mathrm{F}}\Delta_{\alpha}N_{\mathrm{F}}^{-}\lambda\ell_{x}(\mathbf{r},\omega), \label{Current_field_Oest}
\end{gathered}
\end{equation}
where we have introduced $\lambda=\left(1- \left[N_{\mathrm{F}}^{+}/N_{\mathrm{F}}^{-}\right]^{2}\right)$ for notational convenience. Substituting Eq.~(\ref{Current_field_Oest}) into Eq.~(\ref{Response_Func2}) yields equations identical in form to the original in-plane configuration Eq.~(\ref{linresponseLLGInPlane}), and will preserve the $p$-wave symmetry of the coupling matrix element. The only effect is to renormalise the magnon energy
\begin{equation}
    \bar{\omega}_{\mathbf{q}}^{\alpha}=\omega_{\mathbf{q}}^{\alpha} + \frac{2\pi e v_{\mathrm{F}}\mu_{\alpha}\Delta_{\alpha}N_{\mathrm{F}}^{-} \lambda}{c},
\end{equation} 
and exchange coupling
\begin{equation}
\begin{aligned}
\Delta_{\alpha}^{\prime} &= \Delta_{\alpha}+ \frac{2\pi \rho_{s} e v_{\mathrm{F}}\mu_{\alpha}N_{\mathrm{F}}^{+}}{c N_{\mathrm{F}}^{-}}.
\end{aligned}
\end{equation}
For the model parameters used in the main part of the paper, and approximating $\mu_{\alpha} \approx \mu_{\mathrm{B}}$ where $\mu_{\mathrm{B}}=e\hbar/2m_{e}c$ is the Bohr magneton \footnote{The strength of the magnetic moment undergoes Bogoliubov enhancement in a nearly identical manner to the exchange couplings in Eq.~(\ref{BogoliubovEnchancement}), but for the purposes of the order of magnitude estimations here, the Bogoliubov enhancement is on the order of $10^{0}$ and doesn't appreciably affect the magnitude of the renormalisations.}, we find the magnitude of the renormalisation of the magnon energy as
\begin{equation}
\begin{gathered}
    \delta\omega_{\mathbf{q}}^{\alpha}  
    \approx -10^{-4} ~ \mathrm{meV},
    \end{gathered}
\end{equation}
where $\delta\omega_{\mathbf{q}}^{\alpha} =\bar\omega_{\mathbf{q}}^{\alpha}-\omega_{\mathbf{q}}^{\alpha}$. Similarly, we find the magnitude of the renormalisation of the exchange coupling as  
\begin{equation}
\begin{gathered}
    \delta\Delta_{\alpha} 
    \approx 10^{-4} ~ \mathrm{meV},
\end{gathered}
\end{equation}
where again $\delta\Delta_{\alpha}=\Delta_{\alpha}^{\prime}-\Delta_{\alpha}$. The $\mathrm{GHz}$ scale of the above renormalisations is consistent with other treatments of hybrid magnon-plasmon modes that consider only the direct electromagnetic coupling mechanism as the coupling which induces hybridisation \cite{dyrdalMagnonplasmonHybridizationMediated2023,costaStronglyCoupledMagnon2023,Bludov_2019}. Furthermore, we have shown that the angular dependence of the coupling matrix element between magnonic and plasmonic sectors remains unchanged with the inclusion of the direct electromagnetic coupling, which only weakly renormalises its magnitude. Thus we can conclude that the emergent Dirac nodes predicted in the main part of the paper should be robust to the direct electromagnetic coupling between the electron current density and magnetic spins. We note however that the emergent Dirac nodes are not afforded true inversion and time-reversal symmetry protection that is present in other Dirac systems, and including higher-order magnon interaction processes (such as magnon-magnon scattering) could gap out the Dirac point in the hybrid spectrum. Estimations for the size of the contribution from higher-order magnon processes are outside the scope of this paper.
\section{Microscopic derivation of Magnon spectrum and effective exchange interaction} \label{MICMagnons}
In this Appendix we derive the magnon dispersion from a minimal two-sublattice Heisenberg Hamiltonian and the effective exchange coupling of this magnet with electrons. For a two-sublattice Ferrimagnet there are two inequivalent spin sites that are anti-ferromagnetically coupled. Starting with the magnon dispersion, the Heisenberg Hamiltonian of such a system is given as
\begin{equation}
    \mathcal{H}_{s}= J \sum_{\langle ij \rangle} \mathbf{S}_{i,A}\cdot\mathbf{S}_{j,B},
\end{equation}
where $|\mathbf{S}_{A}| \neq |\mathbf{S}_{B}|$. To diagonalise this Hamiltonian we can rotate one of the spin-sublattice sites by a $\pi$ rotation, such that the ground-state appears Ferromagnetic \cite{Altland_Simons_2010}. The tradeoff in doing so is introducing particle non-conserving terms like $\hat{b}\hat{b}$. After performing this rotation and doing a Holstein-Primakoff transformation \cite{PhysRev.58.1098}, we arrive at the intermediate form of the Hamiltonian as 
\begin{equation}
    \mathcal{H}_{s}= \sum_{\mathbf{q}} \Phi_{\mathbf{q}}^{\dagger} H_{\mathbf{q}} \Phi_{\mathbf{q},}
\end{equation}
here $\Phi_{\mathbf{q}} = (a_{\mathbf{q}},b_{-\mathbf{q}}^{\dagger})$ is a spinor of magnon creation/annihilation operators on each spin-subblatice site in reciprocal space, and the Bloch Hamiltonian takes the form
\begin{equation}
    H_{\mathbf{q}}=4J\begin{pmatrix}
        S_{\mathrm{B}} & \sqrt{S_{\mathrm{A}}S_{\mathrm{B}}} \gamma_{\mathbf{q}} \\
        \sqrt{S_{\mathrm{A}}S_{\mathrm{B}}}\gamma_{\mathbf{q}} & S_{\mathrm{A}}
    \end{pmatrix},
\end{equation}
where $\gamma_{\mathbf{q}}=\frac{1}{2}\left[\cos(q_{x}/2)+\cos(q_{y}/2)\right]$ is the geometric factor of the lattice, which we have chosen to be a square lattice for simplicity. To diagonalise the Hamiltonian above, we perform a Bogoliubov transformation of the operators as 
\begin{equation}
    \begin{gathered}
    a_{\mathbf{q}} = u_{\mathbf{q}} \xi_{\mathbf{q},\mathrm{R}} + v_{\mathbf{q}} \xi_{\mathbf{q},\mathrm{L}}^{\dagger}, \\
    b_{\mathbf{q}} = u_{\mathbf{q}}\xi_{\mathbf{q},\mathrm{L}} + v_{\mathbf{q}}\xi_{\mathbf{q},\mathrm{R}}^{\dagger}.
    \end{gathered} \label{bogtrans}
\end{equation}
The parameterisation we choose is $u_{\mathbf{k}}=\cosh(\theta_{\mathbf{k}})$ and $v_{\mathbf{k}}=\sinh(\theta_{\mathbf{k}})$ with $\tanh(2\theta_{\mathbf{k}}) = 2\sqrt{S_{\mathrm{A}}S_{\mathrm{B}}}\gamma_{\mathbf{k}}/(S_{\mathrm{A}}+S_{\mathrm{B}}) $, this transformation diagonalises the Hamiltonian as 
\begin{equation}
    H^{\prime}_{\mathbf{q}} = \begin{pmatrix}
        \omega^{\mathrm{R}}_{\mathbf{q}} & 0 \\
        0 & - \omega^{\mathrm{L}}_{\mathbf{q}},
    \end{pmatrix},
\end{equation}
where 
\begin{equation}
    \omega^{\mathrm{R}/\mathrm{L}}_{\mathbf{q}} = \sqrt{J^{2}(S_{\mathrm{A}}+S_{\mathrm{B}})^{2}-J^{2}S_{\mathrm{A}}S_{\mathrm{B}}|\gamma_{\mathbf{q}}|^{2}} \pm J(S_{\mathrm{A}}-S_{\mathrm{B}}). \label{SpinSpectrum}
\end{equation}
The Hamiltonian that describes the exchange interaction between electron spin $\boldsymbol{\sigma}=(\sigma_{x},\sigma_{y})$ and magnetic sublattice spin is given as
\begin{equation}
    \mathcal{H}_{e-s} = \sum_{i} \Delta_{\mathrm{A}} \boldsymbol{\sigma}\cdot \mathbf{S}_{i,\mathrm{A}} + \Delta_{\mathrm{B}} \boldsymbol{\sigma}\cdot \mathbf{S}_{i,\mathrm{B}}, 
\end{equation}
where $\Delta_{\mathrm{A}/\mathrm{B}}$ are the sublattice onsite exchange energies. Upon performing the Holstein-Primakoff transformation and Fourier transforming, takes the form
\begin{equation}
    \begin{aligned}
    \mathcal{H}_{e-s} &=\\ 
    \sum_{\mathbf{q}}\sqrt{2S_{\mathrm{A}}}&\Delta_{\mathrm{A}}(\sigma^{-}a_{\mathbf{q}}+\sigma^{+}a_{\mathbf{q}}^{\dagger})+\sqrt{2S_{\mathrm{B}}}\Delta_{\mathrm{B}}(\sigma^{-}b_{\mathbf{q}}^{\dagger}+\sigma^{+}b_{\mathbf{q}}),
    \end{aligned}
\end{equation}
where $\sigma^{\pm}=\frac{1}{2}(\sigma_{x}\pm i\sigma_{y})$. To get the final form of the exchange interactions, we need to perform the Bogoliubov transform used to diagonalise the Heisenberg Hamiltonian, Eq.~\ref{bogtrans}. Upon doing so, the final form of the Hamiltonian is easily seen to be
\begin{equation}
    \sum_{\mathbf{q}}\Delta_{\mathrm{R}}(\sigma^{-}\xi_{\mathbf{q},\mathrm{R}}+\sigma^{+}\xi_{\mathbf{q},\mathrm{R}}^{\dagger}) + \Delta_{\mathrm{L}}(\sigma^{-}\xi_{\mathbf{q},\mathrm{L}}^{\dagger}+\sigma^{+}\xi_{\mathbf{q},\mathrm{L}}),
\end{equation}
where
\begin{equation}
\begin{gathered}
\Delta^{\mathrm{R}} = (\sqrt{2S_{\mathrm{A}}}u_{\mathbf{q}}\Delta_{\mathrm{A}}+\sqrt{2S_{\mathrm{B}}}v_{\mathbf{q}}\Delta_{\mathrm{B}}),\\
\Delta^{\mathrm{L}} = (\sqrt{2S_{\mathrm{A}}}v_{\mathbf{q}}\Delta_{\mathrm{A}}+\sqrt{2S_{\mathrm{B}}}u_{\mathbf{q}}\Delta_{\mathrm{B}}), \label{BogoliubovEnchancement}
\end{gathered} 
\end{equation} 
are the effective exchange energies. In Section.~\ref{SecIII} of the main part of the paper, the quoted effective exchange rates of $\Delta_{\mathrm{R}}=70 ~ \mathrm{meV}$ and $\Delta_{\mathrm{L}}=50~\mathrm{meV}$ were calculated with the following set of relevant material parameters: $\mathrm{S}_{\mathrm{A}/\mathrm{B}}=\frac{3}{2}/\frac{1}{2}$ and $\Delta_{\mathrm{A}/\mathrm{B}}=30/10 ~ \mathrm{meV}$. The Bogoliubov enhancement parameters defined above were approximated by their values at $\mathbf{q}= 0$, which is reasonable as the crossing between the magnon and plasmon spectrum happen in the proximity of the $\Gamma$ point.
\section{Derivation of linearised Landau-Lifshitz-Gilbert equations}\label{LLG_eq}
This section derives the linearised LLG equations describing the two magnon chiralities presented in the main text. Consider a two sublattice ferrimagnet, the dynamic of each sublattice magnetic moment follow a set of coupled Landau-Lifshitz-Gilbert equations given as 
\begin{equation}
    \rho_{\mathrm{S}} \partial_{t}\mathbf{n}^{\alpha}(\mathbf{r},t) = \left[\mathbf{B}^{\alpha}_{\mathrm{Eff}} \times \mathbf{n}^{\alpha}(\mathbf{r},t) \right],
\end{equation}
here we gave assumed that the number density of magnetic spins of each sublattice are equal. The effective magnetic field is given as the energy conjugate of the potential energy $\mathbf{B}_{\mathrm{Eff}}^{\alpha}=-\delta U_{\mathbf{n}}/\delta \mathbf{n}^{\alpha} $ which is given as
\begin{widetext}
\begin{equation}
    U[\mathbf{n}_{\mathrm{A}},\mathbf{n}_{\mathrm{B}}] = \int d\mathbf{r} \left\{ \rho_{\mathrm{S}}\left[ J \mathrm{a}^{2}\ \nabla \mathbf{n}_{\mathrm{A}} \cdot \nabla \mathbf{n}_{\mathrm{B}}- \frac{\delta_{s}}{2}((\mathbf{n}_{\mathrm{A}}^{\parallel})^{2}+(\mathbf{n}_{\mathrm{B}}^{\parallel})^{2}) +  2J\ \mathbf{n}_{\mathrm{A}} \cdot \mathbf{n}_{\mathrm{B}}\right] +  \Delta_{\mathrm{A}} \mathbf{n}_{\mathrm{A}}\cdot\mathbf{s} +  \Delta_\mathrm{B} \mathbf{n}_{\mathrm{B}}\cdot\mathbf{s}  \right\},
\end{equation}
\end{widetext}
where $\mathbf{s}$ is the spin density as introduced in the main text. Upon calculating the effective magnetic fields, we get the following coupled LLG equations
\begin{widetext}
\begin{equation}
\begin{gathered}
    \rho_{\mathrm{S}}\partial_{t}\mathbf{n}_{\mathrm{A}} = \rho_{\mathrm{S}}J \mathrm{a}^{2}\left[ \mathbf{n}_{\mathrm{A}} \times \nabla^{2}\mathbf{n}_{B} \right]+ \rho_{\mathrm{S}} 2J\left[ \mathbf{n}_{\mathrm{A}} \times \mathbf{n}_{\mathrm{B}} \right]- \rho_{\mathrm{S}}\delta_{s}\left[\mathbf{n}_{\mathrm{A}} \times \boldsymbol{\ell}_{\mathrm{A}} \right] +\Delta_{\mathrm{A}} \mathbf{n}_{A}\times \mathbf{s}, \\
    \rho_{\mathrm{S}}\partial_{t}\mathbf{n}_{\mathrm{B}} = \rho_{\mathrm{S}}J \mathrm{a}^{2}\left[ \mathbf{n}_{\mathrm{B}} \times \nabla^{2}\mathbf{n}_{\mathrm{A}} \right]+ \rho_{\mathrm{S}}2 J\left[ \mathbf{n}_{\mathrm{B}} \times \mathbf{n}_{\mathrm{A}} \right]- \rho_{\mathrm{S}}\delta_{s}\left[\mathbf{n}_{\mathrm{B}} \times \boldsymbol{\ell}_{\mathrm{B}} \right] +\Delta_\mathrm{B}  \mathbf{n}_{B}\times \mathbf{s}.
\end{gathered}
\end{equation}
\end{widetext}
Now we introduce the following order parameters, we linearise as $\mathbf{n}_{A}=S_{\mathrm{A}}(\mathbf{e}_{z} + \boldsymbol{\ell}_{A})$ and $\mathbf{n}_{B}=S_{\mathrm{B}}(-\mathbf{e}_{z}+\boldsymbol{\ell}_{B})$ which yields, upon rearrangeing, the following linearised LLG equations
\begin{widetext}
    \begin{equation}
    \begin{gathered}
    \left[\partial_{t}\boldsymbol{\ell}_{\mathrm{A}}\times \mathbf{e}_{z}\right] - S_{\mathrm{B}}J \mathrm{a}^{2} \nabla^{2}\boldsymbol{\ell}_{\mathrm{B}} -  2J S_{\mathrm{B}}  (\boldsymbol{\ell}_{\mathrm{A}}+\boldsymbol{\ell}_{B}) +\delta_{s}S_{\mathrm{A}}\boldsymbol{\ell}_{\mathrm{A}}  =\frac{\Delta_\mathrm{A}}{\rho_{\mathrm{S}}} \mathbf{s}, \\
    \left[\partial_{t}\boldsymbol{\ell}_{\mathrm{B}}\times \mathbf{e}_{z}\right] + S_{\mathrm{A}}J \mathrm{a}^{2}\nabla^{2}\boldsymbol{\ell}_{\mathrm{A}} + 2JS_{\mathrm{A}}(\boldsymbol{\ell}_{\mathrm{A}}+\boldsymbol{\ell}_{B})- \delta_{s}S_{\mathrm{B}} \boldsymbol{\ell}_{\mathrm{B}}  =-\frac{\Delta_\mathrm{B}}{\rho_{\mathrm{S}}}  \mathbf{s}.
    \end{gathered}
    \end{equation}
\end{widetext}
To de-couple these two equations, we can formulate the set of coupled equations as a matrix equation with a source term generated by the coupling with electron spin density written as $(\hat{R}(\omega)+\hat{H}(\hat{\mathbf{p}}))\ket{\boldsymbol{\ell}}=\frac{\Delta}{\rho_{S}}\ket{\mathbf{s}}$ where we represent the kets in the cartesian basis as $\ket{\boldsymbol{\ell}} \rightarrow (\ell_{\mathrm{A}}^{x},\ell_{\mathrm{A}}^{y},\ell_{\mathrm{B}}^{x},\ell_{\mathrm{B}}^{y})$, $\ket{\mathbf{s}} \rightarrow (s^{x},s^{y},s^{x},s^{y})$ and
\begin{widetext}
    \begin{equation}
    \begin{gathered}
        \hat{R} = \begin{pmatrix}
            0 & -i \omega & 0 & 0 \\
            i\omega &0 &0 &0 \\
            0 & 0 & 0 & -i\omega \\
            0 & 0 & i\omega & 0
        \end{pmatrix}, \quad \hat{H} = \begin{pmatrix}
            S_{\mathrm{A}}\delta_{s}- 2JS_{\mathrm{B}} & 0 & J\mathrm{a}^{2}S_{\mathrm{B}} \hat{p}_{x}^{2} - 2JS_{\mathrm{B}} & 0 \\
            0 & S_{\mathrm{A}}\delta_{s}- 2JS_{\mathrm{B}} & 0 & J\mathrm{a}^{2}S_{\mathrm{B}} \hat{p}_{y}^{2} - 2JS_{\mathrm{B}} \\
            -J\mathrm{a}^{2}S_{\mathrm{A}}\hat{p}_{x}^{2}+2JS_{\mathrm{A}} & 0 & -S_{\mathrm{B}}\delta_{s}+2JS_{\mathrm{A}}& 0 \\
            0 & -J\mathrm{a}^{2}S_{\mathrm{A}}\hat{p}_{y}^{2}+2JS_{\mathrm{A}} & 0 & -S_{\mathrm{B}}\delta_{s}+2JS_{\mathrm{A}}
        \end{pmatrix}.
        \end{gathered}
    \end{equation}
\end{widetext}
The momentum dependent matrix, $\hat{H}$ is diagonalizable, and thus its Jordan normal form is diagonal with its eigenvalues on the diagonal. Assuming $\delta_{s}=0$ for simplicities sake, we diagonalise via the similarity transformation $\hat{T}^{-1}\hat{H}(\mathbf{p})\hat{T} =\hat{D}$ where $\hat{T}$ is the transition matrix who's columns are the eigenvectors of $\hat{H}$, yields the diagonal matrix 
\begin{equation}
    \hat{D}= \begin{pmatrix}
        \omega^{R}_{\mathbf{p}} & 0 & 0 &0 \\
        0 & \omega^{R}_{\mathbf{p}} & 0 & 0 \\
        0 & 0 & -\omega^{L}_{\mathbf{p}} & 0 \\
        0 & 0 & 0 & -\omega^{L}_{\mathbf{p}}
    \end{pmatrix},
\end{equation}
where $\omega^{\mathrm{R}/\mathrm{L}}_\mathbf{p}$ is the same spectrum derived from the Heisenberg Hamiltonian, Eq.~(\ref{SpinSpectrum}). The form of the diagonal matrix is exactly identical to the diagonalised Heisenberg Hamiltonian after performing a Bogoliubov transformation, indicating that the diagonalisation above is equivalent to the Bogoliubov transformation. Furthermore because this matrix is diagonalised, it is easy to see that we can present the equations as
\begin{equation}
    \hat{R}(\omega) \hat{T}^{-1}\ket{\boldsymbol{\ell}} + \hat{D}(\mathbf{p})\hat{T}^{-1}\ket{\boldsymbol{\ell}} = \frac{\Delta}{\rho_{S}}\hat{T}^{-1}\ket{\boldsymbol{s}},
\end{equation}
which can easily be seen to take the form
\begin{equation}
\begin{gathered}
    \rho_\mathrm{s}\left(\partial_{t}\left[ \boldsymbol{\ell}^{R}(\mathbf{r}, t)\times \vec{e}_{z}\right]-\omega^{R}_{\boldsymbol{q}}\boldsymbol{\ell}^{R}(\mathbf{r}, t)\right)= \Delta_{R} \vec{s}(\vec{r},t), \\
      \rho_\mathrm{s}\left(\partial_{t}\left[\vec{e}_{z}\times \boldsymbol{\ell}^{L}(\mathbf{r}, t)\right]-\omega^{L}_{\boldsymbol{q}}\boldsymbol{\ell}^{L}(\mathbf{r}, t)\right)= \Delta_{L} \vec{s}(\vec{r},t),
\end{gathered}
\end{equation}
where now $\boldsymbol{\ell}^{R} = \sum_{j}(T^{-1}_{1j}\ket{\boldsymbol{\ell}}_{j},T^{-1}_{2j}\ket{\boldsymbol{\ell}}_{j})$ and $\boldsymbol{\ell}^{L} = \sum_{j}(T^{-1}_{3j}\ket{\boldsymbol{\ell}}_{j},T^{-1}_{4j}\ket{\boldsymbol{\ell}}_{j})$

\bibliography{Hybrid_modes_R1}
\end{document}